\documentclass[onecolumn, a4paper, oneside, titlepage]{article}

\pdfoutput=1

\usepackage{lineno}					
\usepackage{amsmath}				
\usepackage{amsfonts}				
\usepackage{setspace}				
\usepackage[]{natbib}	 	
\usepackage{graphicx}				
\usepackage{siunitx}				

\usepackage[superscript]{cite}

\usepackage{tcolorbox}
\tcbuselibrary{breakable}

\usepackage[%
	lmargin=2.5cm,%
	rmargin=2.5cm,%
	tmargin=3cm,%
	bmargin=3cm%
]{geometry}%

\newcommand*\patchenviroforlineno[1]{%
	\expandafter\let\csname old#1\expandafter\endcsname\csname #1\endcsname %
	\expandafter\let\csname oldend#1\expandafter\endcsname\csname end#1\endcsname %
	\renewenvironment{#1}%
		{\linenomath\csname old#1\endcsname}%
		{\csname oldend#1\endcsname\endlinenomath}%
}%

\newcommand*\patchbothenviroforlineno[1]{%
	\patchenviroforlineno{#1}%
	\patchenviroforlineno{#1*}
}%

\AtBeginDocument{%
	\patchbothenviroforlineno{equation}%
	\patchbothenviroforlineno{align}%
	\patchbothenviroforlineno{flalign}%
	\patchbothenviroforlineno{alignat}%
	\patchbothenviroforlineno{gather}%
	\patchbothenviroforlineno{multline}%
}%

\begin{document}
\begin{titlepage}
\vspace{\stretch{1}}
\begin{center}
{ \huge \bfseries %
Eco-evolutionary feedbacks --- theoretical models and perspectives
} \\ %

\vspace{\stretch{0.25}}

{ \large %
Lynn Govaert$^{1,2,3,*}$, Emanuel A. Fronhofer$^{4,*,+}$, S\'ebastien Lion$^{5}$, Christophe Eizaguirre$^{6}$, Dries Bonte$^{7}$, Martijn Egas$^{8}$, Andrew P. Hendry$^{9}$, Ayana De Brito Martins$^{10}$, Carlos J. Meli\'an$^{10}$, Joost A. M. Raeymaekers$^{11}$, Irja I. Ratikainen$^{12,13}$, Bernt-Erik Saether$^{12}$, Jennifer A. Schweitzer$^{14}$ and Blake Matthews$^{2}$
} \\ %
\end{center}

\vspace{\stretch{0.1}}

\begin{enumerate}
\itemsep-1.5em	
\item[1] Laboratory of Aquatic Ecology, Evolution and Conservation, KU Leuven, Deberiotstraat 32, Belgium\\
\item[2] Eawag: Swiss Federal Institute of Aquatic Science and Technology, Department of Aquatic Ecology, \"Uberlandstrasse 133, CH-8600 D\"ubendorf, Switzerland\\
\item[3] Department of Evolutionary Biology and Environmental Studies, University of Zurich, Winterthurerstrasse 190, CH-8057 Z\"urich, Switzerland\\
\item[4] ISEM, Universit\'e de Montpellier, CNRS, IRD, EPHE, Montpellier, France\\
\item[5] Centre d'Ecologie Fonctionnelle et Evolutive, CNRS, Universit\'e de Montpellier, Universit\'e Paul Val\'ery Montpellier 3, IRD, EPHE, Montpellier, France\\
\item[6] Queen Mary University of London, Mile end Road, E14NS, London, United Kingdom\\
\item[7] Ghent University, Dept. Biology, K.L. Ledeganckstraat 35, B-9000 Ghent, Belgium\\
\item[8] Institute for Biodiversity and Ecosystem Dynamics, University of Amsterdam, The Netherlands\\
\item[9] Redpath Museum and Department of Biology, McGill University, Montreal, QC, Canada\\
\item[10] Fish Ecology and Evolution Department, Center for Ecology, Evolution and Biogeochemistry, Eawag, Swiss Federal Institute of Aquatic Science and Technology, Switzerland\\
\item[11] Faculty of Biosciences and Aquaculture, Nord University, N-8049 Bod\o, Norway\\
\item[12] Department of Biology, Centre for Biodiversity Dynamics, Norwegian University of Science and Technology, Trondheim NO-7491, Norway\\
\item[13] Institute of Biodiversity, Animal Health and Comparative Medicine, Graham Kerr Building, University of Glasgow, Glasgow G12 8QQ, United Kingdom\\
\item[14] Department of Ecology and Evolutionary Biology, University of Tennessee, Knoxville, TN 37996, USA\\
\\
\item[*] These authors contributed equally.\\
\item[+] Corresponding author. Orcid ID: 0000-0002-2219-784X
\end{enumerate}

\vspace{\stretch{0.25}}
\begin{flushright}
\textbf{Correspondence Details}\\
Emanuel A. Fronhofer\\
Institut des Sciences de l'Evolution de Montpellier, UMR5554\\
Universit\'e de Montpellier, CC065, Place E. Bataillon, 34095 Montpellier Cedex 5, France\\
phone: +33 (0) 4 67 14 31 82\\
email: emanuel.fronhofer@umontpellier.fr
\end{flushright}
\end{titlepage}

\modulolinenumbers[2]
\doublespacing
\begin{abstract} 
\begin{enumerate}
	\item Theoretical models pertaining to feedbacks between ecological and evolutionary processes are prevalent in multiple biological fields. An integrative overview is currently lacking, due to little crosstalk between the fields and the use of different methodological approaches. 
	\item Here we review a wide range of models of eco-evolutionary feedbacks and highlight their underlying assumptions. We discuss models where feedbacks occur both within and between hierarchical levels of ecosystems, including populations, communities, and abiotic environments, and consider feedbacks across spatial scales.
	\item Identifying the commonalities among feedback models, and the underlying assumptions, helps us better understand the mechanistic basis of eco-evolutionary feedbacks. Eco-evolutionary feedbacks can be readily modelled by coupling demographic and evolutionary formalisms. We provide an overview of these approaches and suggest future integrative modelling avenues.
	\item Our overview highlights that eco-evolutionary feedbacks have been incorporated in theoretical work for nearly a century. Yet, this work does not always include the notion of rapid evolution or concurrent ecological and evolutionary time scales. We discuss the importance of density- and frequency-dependent selection for feedbacks, as well as the importance of dispersal as a central linking trait between ecology and evolution in a spatial context.
\end{enumerate}
\end{abstract}

\section{Introduction}
Feedbacks are relevant to many biological systems and are central to ecology and evolutionary biology \citep{Robertson1991}. While ecology aims to understand the interactions between individuals and their environment, evolution refers to changes in allele frequencies over time. In the past both fields have, to a large extent, been studied in isolation. Evolutionary ecology \citep[e.g.][]{Roughgarden1979} is a notable exception, where links between ecology and evolution are key to empirical and theoretical research.

One of the pioneering studies on feedbacks between ecology and evolution dates back to Pimentel's work on `genetic feedback' \citep{Pimentel1961}. In this feedback, frequencies and densities of different genotypes in a host population shift the overall population density. This changes selection on the host and consequently shifts genotype frequencies. Another early feedback concept of great importance is density-dependent selection \citep{Chitty1967} where the strength of selection changes due to changing population densities, and vice versa \citep[crowding; see also][]{Clarke1972, Travis2013b}.

\begin{figure}[h!]
	\centering
	\includegraphics[width=70mm]{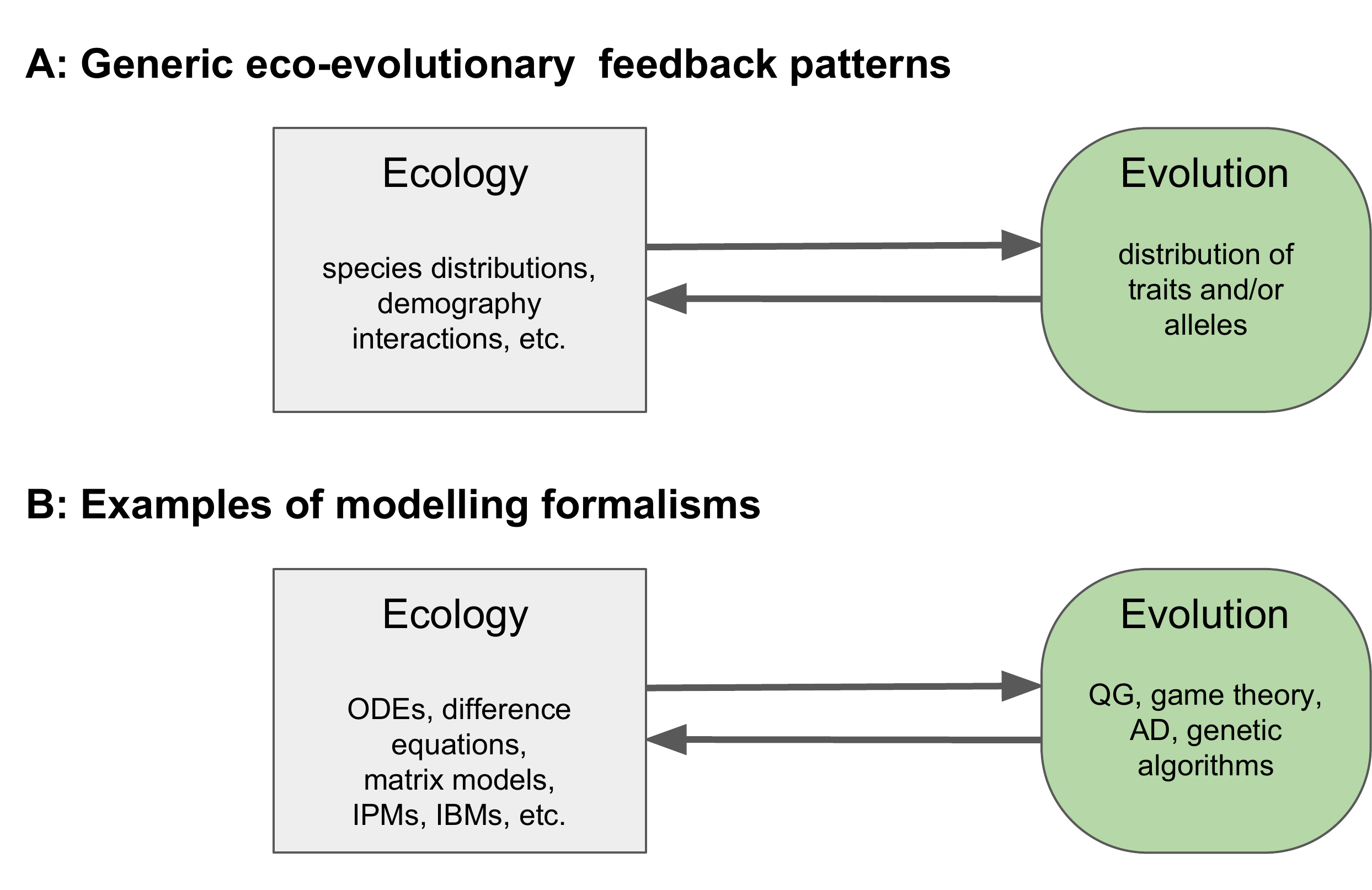}
	\caption{Eco-evolutionary feedbacks (EEF). (A) Generic representation of feedbacks between ecology and evolution implying that the effect of an ecological property (e.g., demography) can be traced to evolutionary change (e.g., shift in allele frequencies; eco-to-evo) and back again to an ecological property (evo-to-eco), or vice versa. (B) Examples of demographic (ecological) and evolutionary modelling formalisms that can be coupled to analyse EEFs. Box types and colours will be used throughout the text to imply ecological or evolutionary aspects, respectively. For a detailed explanation of abbreviations, see Box~1.}
	\label{Fig:introduction}
\end{figure}

In recent years, the recognition that evolution can be rapid and occur on similar timescales as ecology \citep{Hendry1999, Hairston2005} has prompted research at the interface between the two disciplines \citep[often termed `eco-evolutionary dynamics';][]{Hendry2017} and renewed the interest in feedbacks \citep[`eco-evolutionary feedbacks' (EEF); see Fig.~\ref{Fig:introduction}A;][]{Ferriere2004, Post2009}. 

Contemporary theory about EEFs builds on many of the same fundamental ideas established by \citet{Pimentel1961} and \citet{Chitty1967} and feedbacks remain central to the development of evolutionary ecology theory \citep[for recent overview see][]{McPeek2017, Lion2018}. Such feedbacks have been found to generate spatial variation in biotic interactions \citep[geographic mosaic of coevolution;][]{Thompson2005}, impact population regulation and community dynamics \citep[e.g.,]{Abrams1997, Patel2018}, and even promote species coexistence \citep{Kremer2017}, to name but a few examples. Besides theoretical work, empirical and especially experimental tests of eco-evolutionary dynamics and feedbacks have increased recently \citep[e.g.,][]{Yoshida2003, Becks2010, Turcotte2011, Brunner2017}. 

Given the centrality of feedbacks to ecology and evolution, here we provide an overview of theoretical work that includes EEFs \citep[for a comprehensive overview of empirical work see][]{Hendry2017}. Currently, the relevant theory varies in methodological approaches (e.g., quantitative genetics, adaptive dynamics) and thematic subdisciplines (e.g., evolutionary rescue, niche construction) with mostly subtle, and at times semantic, distinctions between them \citep{Matthews2014} which makes concerted progress unnecessarily tedious. In an attempt to bridge these boundaries we organize our non-exhaustive overview around two axes of biological complexity: community (from single to multi-species models) and spatial complexity (from non-spatial to spatially explicit models). Our aims are to summarize available formalisms used to study EEFs theoretically, to highlight their underlying assumptions and to give an overview of existing theoretical work. We will use our overview to flesh out the generic feedback loop shown in Fig.~\ref{Fig:introduction}A and to suggest a more mechanistic representation. 

\section{Formalisms used for modelling EEFs}
Theoreticians use a variety of demographic models to study the interplay between ecology and evolution, including classical ordinary differential equation models (ODEs, e.g., Lotka-Volterra equations, for explanations and abbreviations of recurring terms see Box~1), structured models (matrix models, physiologically structured population models, integral projection models), or stochastic agent-based models. By introducing genetic variation (via standing genetic variation and/or mutations) in one or several populations, the models can capture EEFs (Fig.~\ref{Fig:introduction}B). Such models can often not be easily analysed mathematically, except in very simple and potentially unrealistic cases. Therefore, various formalisms, such as adaptive dynamics (AD) and quantitative genetics (QG), have been developed to further our understanding of EEFs \citep[reviewed in][]{Lion2018}. 

Models using AD rely on a separation of time scales between ecological and evolutionary dynamics. Specifically, these models assume that mutations are so rare that the ecological community is always on its attractor, so that the evolutionary dynamics take the form of a temporal sequence of allele substitutions (i.e., mutation-limited evolution). The success of a mutant allele is then measured by its invasion fitness \citep{Metz1992, Geritz1998}. The separation of time scales between ecology and evolution, however, does not mean that there is no EEF. The feedback is materialised by the fact that the invasion fitness of a mutant allele depends on the ecological conditions created by the resident community. In fact, the concept of a `feedback loop' between ecology and evolution has been central in the development of AD \citep{Ferriere2012}.

QG models, by contrast, start from a different perspective and explicitly consider evolution resulting from existing genetic variation. For a given quantitative trait, these models track the dynamics of different moments of the trait distribution \citep[mean, variance, etc; distribution shapes can central to eco-evolutionary dynamics:][]{Chevin2017}. Often, additional assumptions have to be made, to allow for simplifications. Many QG models assume that the trait distribution is Gaussian and tightly clustered around the mean (small variance or weak selection approximation). In that case, it becomes possible to approximate the ecological dynamics of the focal population as if all individuals had the mean trait value, and to understand the change in mean trait in relation to a selection gradient, where the selection gradient itself depends on the ecological dynamics \citep[e.g.,][]{Abrams1997, Luo2013, Lion2018}. This allows the coupling of ecology and evolution, similarly to AD, with the difference that ecological dynamics do not have to be at equilibrium \citep[no separation of time scales; see][for the impact of environmental variation]{Lande2007, Lande2009}. Therefore, QG models can focus on short-term dynamics, which makes them potentially more applicable to experiments or field studies where rapid evolution is a key process.

On the demographic (ecological) side, ODEs, matrix models (e.g., integral projection models --- IPMs) and individual-based models (IBMs) have used the AD and QG approach to investigate EEFs. ODEs and matrix models have been long used to study simultaneous change between ecological (e.g., population size) and evolutionary parameters (e.g. strength of selection), without explicitly using the term EEF \citep[see e.g.,][]{Caswell2006}. IPMs have recently been explicitly coupled to QG and AD \citep{Rees2016}. IBMs have the advantage of including demography and evolutionary change, which emerges via selection pressures from the ecological setting \citep[see also genetic algorithms;][]{Fraser1957}. IBMs lend themselves very easily to the incorporation of complexities such as stochasticity, spatial structure and kin competition \citep[e.g.][]{Poethke2007}. The downside of IBMs is a loss of generality and often also tractability.

\begin{tcolorbox}[breakable=true,title=Box 1: Explanation of terms and abbreviations]
\textbf{Adaptive dynamics (AD)}: AD is a mathematical formalism, that provides a dynamical extension of classical optimisation approaches and evolutionary game theory to include density- and frequency dependence \citep{Diekmann2004, Waxman2005}. This makes eco-evolutionary feedbacks central to AD.

\textbf{Dispersal}: Dispersal is the movement of individuals away from their parents with potential consequences for gene flow \citep{Clobert2012}.

\textbf{Eco-evolutionary feedback (EEF)}: For the purpose of this overview EEFs will imply that we can trace the effect of an ecological property (e.g., demography) to evolutionary change (e.g., shift in allele frequencies; eco-to-evo) and back again to an ecological property (evo-to-eco; Fig.~\ref{Fig:introduction}A), or vice versa \citep[narrow and broad sense feedbacks sensu][]{Hendry2017}. 

\textbf{Evolutionary rescue (ER) and suicide (ES)}: ER is the idea that a population can avoid extinction through rapid adaptation \citep{Gonzalez2013}. By contrast, ES is the process by which evolution drives a population beyond its viability region, eventually causing extinction \citep{Ferriere2000a}.

\textbf{Individual-based model (IBM)}: IBM (also agent-based model, ABMs) are bottom-up models in which a (meta)population or (meta)community is modelled as a number of discrete interacting individuals, in which each individual is characterized by a set of state variables (location, physiological or behavioural traits). The interactions between individuals result in (meta)population- and (meta)community or (meta)foodweb dynamics \citep{Grimm1999, DeAngelis2005}.

\textbf{Integral projection model (IPM)}: IPMs describe the dynamics of a population by projecting its size or trait distribution through time using a kernel distribution that connects individual-level vital rates such as survival, reproduction and development to population-level processes. IPMs can be coupled with AD or QG approaches \citep{Rees2016}.

\textbf{Lotka-Volterra model (LV)}: The LV model (named after Alfred Lotka and Vito Volterra) consists of ODEs describing predator and prey dynamics. Modifications of the basic model include e.g. the Rosenzweig-MacArthur model.

\textbf{Metapopulation and metacommunity}: A metapopulation \textit{sensu lato} is a spatially structured population, connected by dispersal \citep{Hanski1999, Harrison1996}. Similarly, a metacommunity is a spatially structured community, connected by dispersal \citep{Leibold2004}. 

\textbf{Quantitative genetics (QG)}: QG studies the genetic basis of phenotypic variation, with a focus on the dynamics of continuous trait distributions \citep{Lynch1998}.
\end{tcolorbox}

\section{EEFs within populations}
Many theoretical studies have analysed EEFs within a single population in a temporal or spatial setting. In single species non-spatial settings, EEFs are usually considered between changes in population size and changes in heritable traits. In a spatial setting, EEFs can occur between local population size and local trait values, but also among patches between regional (meta)population size and local or regional trait values. In addition, landscape structure (topology, connectivity) might influence local EEFs, but also induce feedbacks on a regional scale. This is because dispersal (demography) and gene flow (population genetics) are intrinsically linked.

\subsection{Feedbacks in single populations}
Feedbacks can occur between population density and trait values, or between the availability of resources and trait values. For example, a quantitative trait subject to density-dependent or frequency-dependent selection (eco-to-evo) can influence population growth rate \citep[evo-to-eco;][]{Lande2007, Engen2013, Travis2013b}. Density- or frequency-dependent selection implies that an individual's fitness is not only determined by its trait value, but also by the population density or by the proportion of certain genotypes \citep{Clarke1972, Travis2013b}. In the case of density-dependent selection, changing population densities shift the selection pressure favouring different genotypes because of differential competitive ability. In turn, changing competitive abilities create varying ecological conditions leading to changes in density \citep{MacArthur1962, Lande2007, Engen2013}.

\begin{figure}[h!]
	\centering
	\includegraphics[width=70mm]{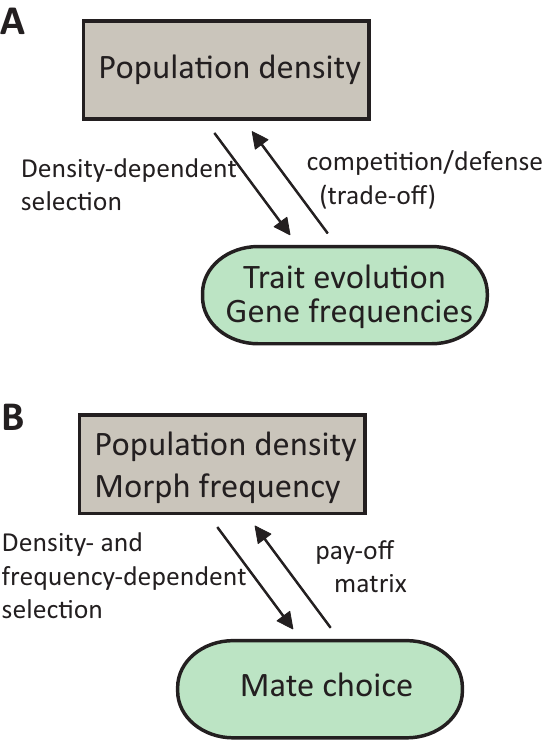}
	\caption{Examples of studies in which feedbacks occur in a single species non-spatial setting. (A) In \citet{Lande2007} and \citet{Lively2012} population density determines the selection pressure, resulting in evolution of some quantitative trait \citep{Lande2007} or shifts in discrete genotype frequencies \citep{Lively2012}. (B) In \citet{Alonzo2001} not only population density but also the frequency of morphs determine mate choice, which in turns determines the outcome of morph frequencies in the next generation influencing the trait mate choice again.}
	\label{Fig:feedback_singleSpec_temp}
\end{figure}

\citet{Lively2012} used a one-locus two-allele genetic system (QG with two types) to illustrate a feedback between population density and allele-frequency change assuming density-dependent selection (Fig.~\ref{Fig:feedback_singleSpec_temp}A). Similarly \citet{Lande2007} and \citet{Engen2013} used QG models linking the evolution of a quantitative trait to population growth, strength of density dependence and environmental stochasticity. These authors found that in a constant environment, evolution will maximize mean fitness and mean relative fitness in the population which may change when population sizes fluctuate \citep{Saether2015}. Technically, the evolutionary response of the population due to a changing environment in these models is described using the phenotypic selection differential (accounting for individual survival and fecundity, but not inheritance) or in terms of the selection gradient \citep{Leon1978, Lande2009}.

The assumption of frequency-dependent selection is particularly relevant in the context of sexual selection and mate choice \citep{Alonzo2001}. Evolutionary game theory can be used to model a population consisting of female and male morphs where female mate preference depends on the total population size (density-dependent selection), but also on female morph frequency (frequency-dependent selection; Fig.~\ref{Fig:feedback_singleSpec_temp}B). This leads to an EEF between population size and morph frequencies via density- and frequency-dependent selection (eco-to-evo) and via fitness differences of the morphs \citep[evo-to-eco; reviewed in][]{Smallegangethisissue}.

Very similar mechanisms have been discussed in the context of the evolution of cooperation \citep[e.g.,][]{Lehtonen2012, Gokhale2016}. For example, ecological conditions, such as resource limitation and variability may select for the evolution of cooperation (eco-to-evo), which can then feed back on demography leading to increased population sizes \citep[``supersaturation'',][]{Fronhofer2011a}.

Finally, a classical EEF over time is often termed evolutionary rescue \citep[ER, see Box~1;][]{Lynch1993, Gomulkiewicz1995, Gonzalez2013}. ER models have either used a QG approach, focusing on the population's capacity to track gradually changing optima in time \citep{Burger1995, Lande1996} or space \citep{Pease1989, Polechova2009, Uecker2014} or a single mutation approach in which a population is exposed to a sudden severe environmental change \citep{Gomulkiewicz1995, Orr2014, Uecker2017}. Interestingly, while ER leads to population persistence, adaptive evolution might also result in evolutionary trapping or suicide \citep[ES,][]{Ferriere2000a, Parvinen2013}. In the latter, trait change drastically degrades population viability leading to extinction \citep{Ferriere2012, Engen2017}. Whether the result is ER or ES, these models demonstrate that EEFs can be of applied relevance to conservation.

In summary, feedbacks over time are usually mediated by intrinsically (density- / frequency dependent selection) or extrinsically (environment) changing selection pressures. The consequences of these feedbacks may be positive (e.g., increased densities and survival) or negative (ES) at the population level.

\subsection{Feedbacks in spatially structured populations}
Spatial models allow for EEFs between local demography or metapopulation conditions and an evolving trait. The feedback can be modified by external properties such as patch dynamics \citep[colonization and extinction rates;][]{Hanski2011} or landscape structure \citep{Kubisch2016, Fronhofer2017}. In models with discrete habitat patches, dispersal is a central trait connecting local patches, and can have important effects on both ecological \citep{Clobert2012} and evolutionary \citep[e.g., can limit or favour local adaptation;][]{Lenormand2002, Raesaenen2008, Nosil2009} processes. The evolution of dispersal likely is the most frequently studied example of an EEF in fragmented landscapes \citep{Legrand2017}.

In a spatial model without dispersal evolution, \citet{Gomulkiewicz1995} show that ER (local adaptation) can be strongly hampered by stochasticity, e.g., as a consequence of low population sizes \citep[see][for another example of spatial ER]{Gomulkiewicz1999}. Interestingly, the probability of rescue can be a non-monotonic function of migration rates \citep{Uecker2014}.

\begin{figure}[h!]
	\centering
	\includegraphics[width=70mm]{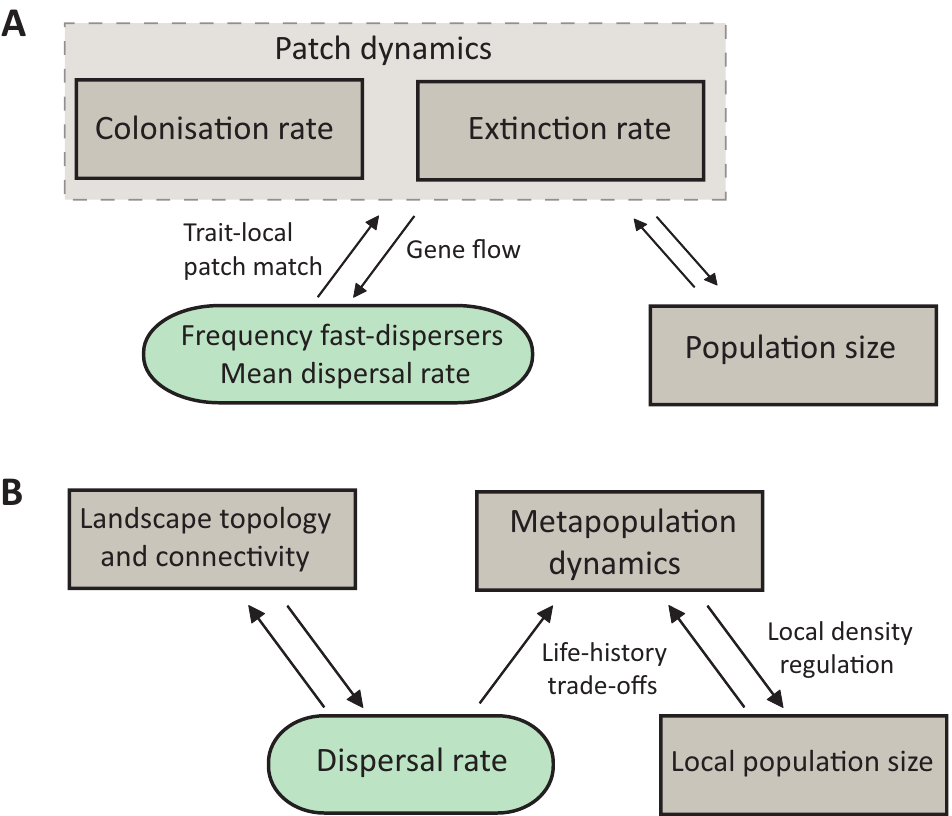}
	\caption{Examples of studies with spatial feedbacks. (A) Study by \citet{Hanski2011a} and \citet{Hanski2011} where patch dynamics driven by colonisation and extinction might influence disperser frequency \citep{Hanski2011} or shifts mean dispersal rate \citep{Hanski2011a}, which in turn influences patch dynamics. (B) Study by \citet{Fronhofer2017} in which landscape topology influences dispersal evolution, which in turn influences colonization probabilities and metapopulation dynamics (occupancy, turnover, genetic structure, global extinction risk).}
	\label{Fig:feedback_singleSpec_spatial}
\end{figure}

If dispersal is allowed to evolve \citep{Ronce2007}, it may be modelled as a discrete trait with dispersing and resident genotypes \citep[e.g.,][]{Hanski2011}, as a quantitative trait \citep{Hanski2011a}, or even as an evolving reaction norm \citep[][for an overview on the genetics of dispersal and how dispersal is incorporated into models see \citealt{Saastamoinen2018}]{Travis1999, Poethke2002}. For example, combining stochastic patch occupancy models with description of mean phenotypic changes in local populations, \citet{Hanski2011} studied an EEF between patch dynamics (colonisation and extinction) and the frequency of a disperser genotype (for details see Fig.~\ref{Fig:feedback_singleSpec_spatial}A).

In spatial models, EEFs can link processes at different spatial scales. For instance, \citet{Poethke2011} show that the selective increase of patch size, e.g., as a conservation measure, can select against dispersal (eco-to-evo) which decreases re-colonization probabilities and can lead to ES (evo-to-eco). In contrast, evolution can rescue populations from extinction in a spatial setting if environmental changes are not too fast \citep{Schiffers2013}. Similarly, in a range expansion context, \citet{Burton2010} and \citet{Fronhofer2015b} showed that the ecological process of a range expansion can select for increased dispersal at range fronts \citep{Travis2002} and may feed back on the distribution of population densities across the range via life-history trade-offs. The importance of landscape structure for EEFs is laid out in \citet{Fronhofer2017} (Fig.~\ref{Fig:feedback_singleSpec_spatial}B).

Taken together, spatial models may consider local adaptation to abiotic conditions as a heritable trait and fix dispersal or may consider dispersal as an evolving trait. Altogether, the studies show that dispersal is an excellent candidate to link ecology (demography from a single population or metapopulation) and evolution, making dispersal central to EEFs.

\section{EEFs involving two species}
In multi-species systems, EEFs can be mediated by intra- and interspecific densities that affect fitness and trait distributions \citep{Travis2013b}. Here, we consider three major categories of two-species interactions: interspecific competition, predator-prey and parasite-host interactions.

\subsection{Interspecific competition}
\begin{figure}[h!]
	\centering
	\includegraphics[width=70mm]{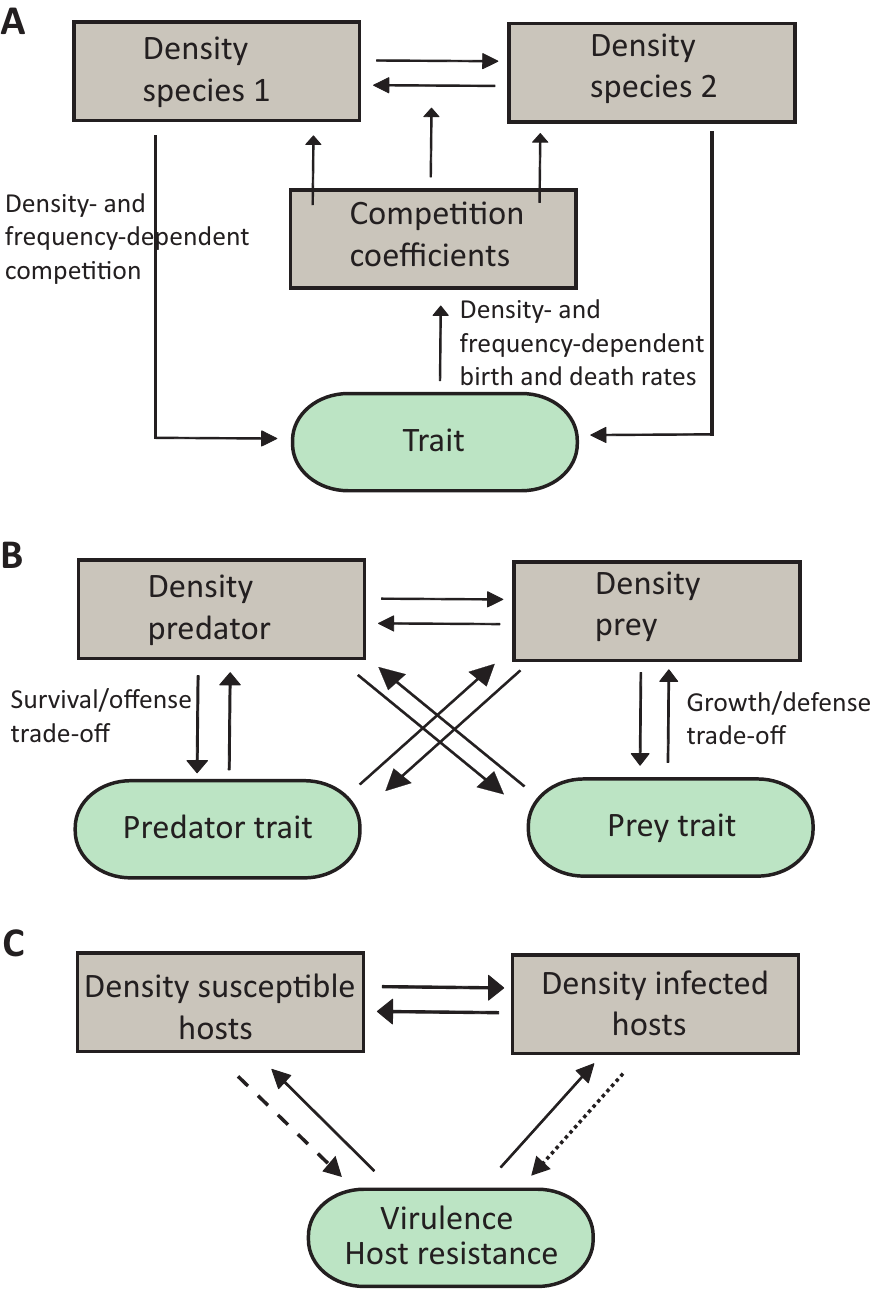}
	\caption{Examples of studies in which feedbacks occur in two-species settings. (A) Study by \citet{Vasseur2011} in which the competition coefficients determining the strength of intra- and interspecific competition are modelled in function of an evolvable trait (growth or defence trait) under density- and frequency-dependent competition. (B) General figure on possible EEFs in predator-prey dynamics \citep[detailed in][]{Cortez2014}. Generally, a trade-off between growth and predator defence is assumed in the prey population, and a trade-off between mortality and offense is assumed in the predator population. Density of the predator and prey can both influence trait evolution in the predator and prey population, which due to the previously described trade-off, determines predator and prey density. (C) General figure on possible feedbacks in host-parasite dynamics \citep[see][]{Luo2013}. In a model of virulence evolution, density of susceptible hosts determines the degree of virulence which feeds back to change the density of susceptible hosts (striped arrow). In a model on host resistance, density of the infected hosts determine the evolution of host resistance (dashed arrow), which in turn determines the density of both susceptible and infected hosts.}
	\label{Fig:feedback_2spec}
\end{figure}

Interspecific competition is a reciprocal interaction for a shared limiting resource \citep{Dhondt1989}, such as food, and competing species can evolve different niches in order to coexist \citep{Brown1956, Abrams1986, Taper1992}. Many studies have shown that competition-induced selection can result in adaptive divergence through ecological character displacement \citep{Brown1956, Slatkin1980, Taper1992, Schluter2000}. In such models, EEFs may occur because competing species exert selection pressures that result in trait evolution (eco-to-evo) that might alter selection pressures on both species (evo-to-eco) \citep[e.g.,][Fig.~\ref{Fig:feedback_2spec}A]{Vasseur2011}.

For example, \citet{Dieckmann1999} used an IBM, in which the evolving trait determines the carrying capacity (competition), and in which individuals survive and die via density- and frequency-dependence giving rise to a feedback between density and trait evolution, resulting in speciation via evolutionary branching. The authors showed that evolution of assortative mating can lead to reproductive isolation, resulting in increased diversity and that non-random mating is a prerequisite for evolutionary branching \citep[see also][]{Thibert-Plante2009}. In a similar model, \citet{Aguilee2013} found that landscape structure highly influences the outcome of diversity resulting from underlying dynamics of competition and assortative mating. The latter study used an IBMs assuming density-dependent resource competition and stronger competition between individuals with similar trait values, inducing frequency-dependent selection and considered traits linked to resource utilization and to mate choice.

In summary, prominent examples of EEFs in two-species competitive systems, focus on character displacement and potentially speciation. These models may include a relatively high level of biological complexity which makes the use of IBMs widespread, with the caveat of limited generality \citep[but see][]{Pennings2008}. Nevertheless, analytical AD models are very well established \citep[see e.g.,][]{Kisdi1999a}.

\subsection{Predator-prey interactions}
EEFs in predator-prey systems imply that predator densities may induce trait evolution in prey defence (eco-to-evo) resulting in consequent shifts in prey and predator densities (evo-to-eco; Fig.~\ref{Fig:feedback_2spec}B). Many studies have found that rapid evolution in prey defence due to shifting predator abundances results in antiphase cycles rather than $\frac{1}{4}$-lag cycles predicted by non-evolutionary models \citep{Yoshida2003, Yoshida2007, Becks2010}. Additionally, feedbacks can stabilize or destabilize predator-prey dynamics depending on genetic variation and trade-off shapes \citep{Abrams1997, Abrams2000, Cortez2010, Cortez2016}. 

Predator-prey dynamics have been extensively studied using models of trait evolution of the prey \citep[e.g.][]{Abrams1997, Cortez2016, McPeek2017}, the predator \citep{Cortez2010}, or both \citep[e.g.][Fig.~\ref{Fig:feedback_2spec}B]{Cortez2014, Velzen2017}. In all three instances EEFs were modelled using either separate equations for the ecological and evolutionary dynamics \citep[e.g.][]{Abrams1997} or QG recursion equations or an approximation of those \citep{Velzen2017}, using an AD approach \citep{Marrow1996} or by using multiclonal LV equations \citep[which are identical to `ecological selection' models][]{Jones2007a, Ellner2011, Yamamichi2011, Cortez2014, Haafke2016}. Including life-history trade-offs between defence and fecundity may lead to recurrent EEFs \citep{Meyer2006, Huang2017}.

Phenotypic plasticity has been found to play an important role in predator-prey EEFs and has been incorporated for example by \citet{Yamamichi2011}, who found that plasticity in prey defence promotes stable population dynamics more than rapid evolutionary responses, although, plasticity was not advantageous in stable environments. The evolution of plasticity has been studied by \citet{Fischer2014a}, who extended an LV model allowing for variation in plasticity among multiple genotypes of prey. The inclusion of such variation in models improved their ability to explain predator-prey dynamics.

Overall, predator-prey EEFs are a classical example of feedbacks involving phase shifts and impacts on stability. These effects are classically modelled with ODEs. Recent work highlights the importance of incorporating both, effects of genetic diversity and phenotypic plasticity to explain community dynamics \citep{Yamamichi2011, Kovach-Orr2013}.

\subsection{Host-parasite interactions}
Just as predators, parasites can impose strong selection pressures on their hosts, resulting in the evolution of defence strategies that can in turn impose selection on parasite life-history traits. This process can lead to complex co-evolutionary dynamics in spatial and non-spatial settings. Due to high mutation rates and genetic diversity of parasite populations, host-parasite interactions are often characterised by overlapping time scales between epidemiological and evolutionary processes. Even when evolution is slower than epidemiology, selection in host-parasite systems is characterised by strong density-dependent feedbacks, where changes in densities affect selection pressures on transmission, virulence and other parasite traits (eco-to-evo), and the resulting trait changes in turn alter the ecological dynamics \citep[evo-to-eco;][Fig.~\ref{Fig:feedback_2spec}C]{Luo2013}.

The study of virulence evolution in parasites and pathogens is a key topic in the theoretical literature involving EEFs. The seminal work of \citet{Anderson1982} showed that pathogen evolution is shaped by the epidemiological dynamics of infectious diseases through the density of susceptible hosts. Since then, a large literature has followed devoted to understanding the effect of EEFs on virulence evolution, and other traits such as host resistance \citep[e.g.][]{Lenski1994, VanBaalen1998, Boots1999a, Frickel2016}. Central to this literature, although sometimes not explicitly so, is the concept of environmental feedbacks \citep[reviewed in][]{Dieckmann2002, Lion2018b}. In particular, much interest has been devoted to cases where the feedback is of sufficiently high dimension to allow for diversification of the pathogen \citep[clusters of low and high virulence;][]{Best2010, Lion2018b}, which in turn can select for evolutionary branching in host resistance \citep{Nuismer2008, Eizaguirre2009, Best2010}.

Most models of host-parasite EEFs use classical epidemiological models (compartment models that include susceptible, infected and potentially recovered individuals) to describe the changes in density or frequency of susceptible and infected hosts. These ODEs then can be coupled with AD, if one is interested in the long-term evolution \citep{Dieckmann2002, Lion2018b}. However, many infectious diseases have high mutation rates or standing genetic variance, resulting in overlapping time scales between ecology and evolution. This has motivated the use of QG or population genetics \citep[e.g.,][]{Day2004, Day2007}. Because AD and QG differ in their time scale assumptions, they allow to investigate how these time scales affect EEFs \citep{Lion2018}. 

Host-parasite interactions have also been a key in understanding spatial EEFs \citep[e.g.,][reviewed in \citealt{Lion2015}]{Boots1999, Boots2004}. These studies have often modelled space as a regular network of sites, in which each site is either empty or contains a single host individual, which can be either susceptible or infected. Such models can easily be analysed using IBMs, but analytical insight are also possible to some extent, using either AD or QG \citep{Lion2016}. Due to the inherent complexity of spatial models, however, we only have a partial understanding of how the feedback between spatial epidemiological dynamics and the evolution of host and parasite traits unfolds in more realistic host-parasite interactions \citep[but see][]{Nuismer2000, Nuismer2003}.

In summary, the host-parasite literature has a long tradition of studying EEFs. Methodological approaches differ depending on the level of complexity, from simple ODEs to IBMs.

\section{EEFs in a community and ecosystem context}
The increasing interest in more complex ecological settings has resulted in a rapid growth of models focusing on communities and ecosystems that could simultaneously incorporate evolutionary dynamics \citep{Brannstrom2012}. Such models extend previous work to include niche construction, plant-soil feedbacks, multiple-species communities and foodwebs. 

\subsection{Feedbacks between organisms and abiotic environments}
\begin{figure}[h!]
	\centering
	\includegraphics[width=70mm]{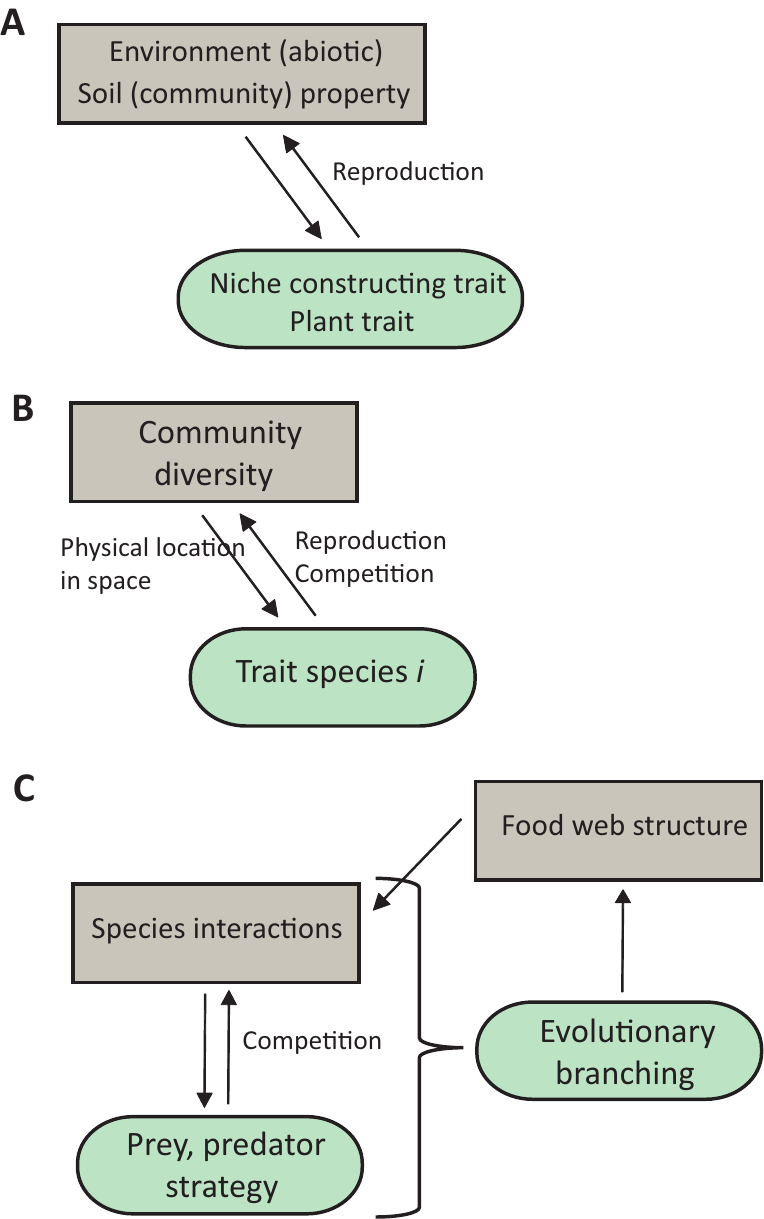}
	\caption{Examples of studies in which feedbacks occur between abiotic and biotic component or in a multi-species settings. (A) General figure of EEFs in niche construction \citep{Lehmann2008, Kylafis2011} and plant-soil feedbacks \citep{Schweitzer2014}. In niche construction the abiotic environment determines the evolution of a trait that modifies this abiotic environment. Similarly, in a plant-soil system, a plant trait can modify the soil, which drives evolution of plant traits. (B) Study by \citet{Martin2016} in which trait values and spatial locations species determine competition, changing local selection pressures, resulting in shifting local and global trait distributions and community diversity. (c) Study by \citet{Ito2006}, in which each species has a separate prey and predator strategy which results in clusters of trophic species arising from changing interactions between species, which in turn continuously change the position, shape and size of occupied areas in phenotypic space and change trophic interactions resulting in further phenotypic evolution and eventually evolutionary branching and the emergence of foodweb structure.}
	\label{Fig:feedback_commEcoSyst}
\end{figure}

EEFs with the environment may involve niche construction \citep{Odling-Smee2003, Lehmann2008, Kylafis2011}, as in plant-soil feedbacks, for example \citep[][Fig.~\ref{Fig:feedback_commEcoSyst}A]{Schweitzer2014, Wareinprep}. Game theory has been used to investigate selection on niche constructing phenotypes \citep[][]{Lehmann2008} where the feedback arises when individuals affect their environment by reproducing (evo-to-eco), hence altering the selection pressure on the population (eco-to-evo). In plant-soil systems, plants might adaptively regulate soil fertility, resulting in positive, self-sustaining nutrient feedbacks that influence evolution. For example, increasing the direct benefit of soil nutrient conditioning to plants has been predicted to increase selection for higher values of soil conditioning traits \citep{Kylafis2008}. Implicit in this model is a genetic link between plants and soils, and subsequent models have shown how genetically based plant-soil feedbacks can evolve along soil gradients \citep{Schweitzer2014}.

In plant-soil systems evolutionary change in plant traits can influence ecological dynamics of soil microbes (evo-to-eco) which in turn can change selection pressures on plant traits (eco-to-evo). This can be investigated using IBMs \citep{Schweitzer2014} or by using an extended version of classical resource-competition models \citep{Eppinga2011}. In this specific model, the decomposition of litter releases nutrients that can be taken up by the plants influencing competitive ability of the plant (eco-to-evo), resulting in different plant genotypes that might grow better. The change in the genetic composition of the plant population can in turn influence the litter pool (evo-to-eco). 

In analogy to negative niche construction \citep{Odling-Smee2003}, the spatial structure of local negative feedbacks can result in changes in local diversity \citep[e.g.,][]{Loeuille2014}. The environment becomes less suitable for the species occupying it (evo-to-eco), which induces a change in selection pressure on the species to evolve toward a more matching trait-environment value (eco-to-evo). 

Overall, plant-soil interactions are good examples of niche construction whereby EEFs can both be modelled and observed in nature. Methods employed reach from formal mathematical approaches to IBMs.

\subsection{Feedbacks within communities}
Theoretical studies on EEFs in multi-species communities are highly valuable to improve our understanding of biodiversity \citep{Patel2018}. Eco-evolutionary analyses have led to new insights into coexistence theory, the maintenance of diversity, as well as the structure and stability of communities \citep{Kremer2017, Patel2018}. Moreover, studies have found that evolution might maintain \citep{Martin2016}, increase \citep[e.g. via speciation or ER][]{Rosenzweig1978, Dieckmann1999, Gomulkiewicz1995} or decrease \citep{Norberg2012, Kremer2013, Gyllenberg2002} phenotypic, species and functional diversity. 

For example, \citet{Martin2016} show that EEFs can maintain phenotypic diversity. The authors combine niche based approaches with neutral theory in a spatially structured IBM where each individual has a location in space and is constrained by a specific trade-off between resource exploitation and competition. Similar individuals experience higher competition resulting in frequency-dependent selection. Competition takes only place between neighboring individuals, changing local selection pressures, which results in local evolutionary shifts in phenotypic traits (eco-to-evo) that shift the global phenotypic trait distribution and influence species differentiation and thus community diversity (evo-to-eco; Fig.~\ref{Fig:feedback_commEcoSyst}B). By contrast, \citet{Norberg2012} found that the eco-evolutionary processes induced by climate change continued to generate species extinctions long after the climate had stabilized, and thus resulted in further diversity loss. These authors used a spatially explicit IBM to predict species responses to climate change in a multi-species context in which they allowed genetic variation and dispersal to jointly influence ecological (competition and species sorting) and evolutionary (adaptation) processes.

In summary, EEFs in communities emerge, because species' traits may affect the community and, vice versa, the community context may affect trait evolution \citep{terHorst2018}. Interestingly, fitness may not only depend on densities, but also on total community biomass, total productivity, or even on species numbers.

\subsection{Feedbacks in food webs}
Eco-evolutionary processes have also been included in food web models \citep{Brannstrom2012, Allhoff2016}. For example, \citet{Ito2006} include evolution via random mutations. Each species has a prey and predator strategy as well as a functional response (as opposed to implicitly present in most food web models). \citet{Ito2006} show that clusters of trophic species arise due to changing interactions between species (eco-to-evo), which in turn continuously change the position, shape and size of occupied areas in phenotypic space and changes trophic interactions (evo-to-eco) resulting in further phenotypic evolution and eventually evolutionary branching (Fig.~\ref{Fig:feedback_commEcoSyst}C). A similar approach has been taking by \citet{Takahashi2013}, in which an IBM was used where each individual is characterized by two evolving traits: foraging and vulnerability traits evolving via mutations. They showed that initial phenotypic divergence in the foraging trait relaxes interference competition (eco-to-evo), which results in the emergence of species clusters, which changes species interactions (trophic levels; evo-to-eco) resulting in further trait divergence in foraging as well as vulnerability (eco-to-evo). Finally, \citet{Andreazzi2018} model antagonistic species networks and explicitly evaluate the effects of EEFs on long-term ecological network stability. The authors showed that EEFs resulted in specific patterns of specialization which led to increases in mean species abundances and to decreases in temporal variation in abundances. 

Overall, eco-evolutionary foodweb models, and especially evolutionary metafoodweb models remain rare. Existing models often include evolution by adding new species from a pool \citep[e.g.,][]{Allhoff2016}, and do not model speciation as a consequence of the intrinsic ecological dynamics of the food web \citep[but see][]{Ito2006}. Likely represent one of the current major challenges in eco-evolutionary modelling \citep{Melian2018}.

\section{Discussion}
Throughout this overview, we found that including EEFs in theoretical models significantly changes our view of well-known patterns emerging from pure ecological or pure evolutionary models \citep[e.g.,][]{Dieckmann2006, Poethke2011}. More specifically, we have identified models that include EEFs, whose underlying formalisms fall into a few categories (Fig.~\ref{Fig:introduction}B). In principle, any modelling framework that couples ecological dynamics (e.g., ODEs, IBMs) with an evolutionary model (e.g., QG, AD or a genetic algorithm) can be useful for studying feedbacks. 

Based on our non-exhaustive overview of theoretical work on EEFs, a few general conclusions emerge: First, density- and frequency-dependent selection are key ingredients for EEFs. In genuinely eco-evolutionary models density- and frequency-dependency are not a priori assumptions, but emerge from ecological settings and trait correlations, for example. Second, EEFs are not new to evolutionary ecology theory --- they are deeply rooted in the theory of many subdisciplines. For instance, the predator-prey and host-parasite literature, speciation literature and evolutionary branching, character displacement, as well as metapopulation modelling or niche construction theory naturally incorporate EEFs. Strikingly, EEFs seem to have been included in (meta)community ecology rather late, which may explain why, at least in this field, the interaction between ecology and evolution seems rather recent \citep[maybe culminating in the recognition that the basic drivers of evolution and community ecology are analogous,][]{Vellend2010}. This disconnection is also visible between single species spatial models that often include evolutionary dynamics, whereas metacommunity models remain mostly ecological. Third, in a spatial setting dispersal is a primary candidate for successful eco-evolutionary linkages, because dispersal is both an ecological process impacting densities and at the same time mediates evolution via gene flow. In addition, it is itself subject to evolution \citep{Ronce2007}. Fourth, EEFs do not necessarily require rapid or contemporary evolution. Of course, contemporary evolution has sparked a lot of interest in EEFs \citep{Hendry2017}, but feedbacks are also possible over longer timescales. Fifth, our short overview of the eco-evolutionary modelling toolbox clearly highlights that the main character of an eco-evolutionary model is the combination of demographic and evolutionary models, regardless of the concrete formalism. 

\begin{figure}[h!]
	\centering
	\includegraphics[width=140mm]{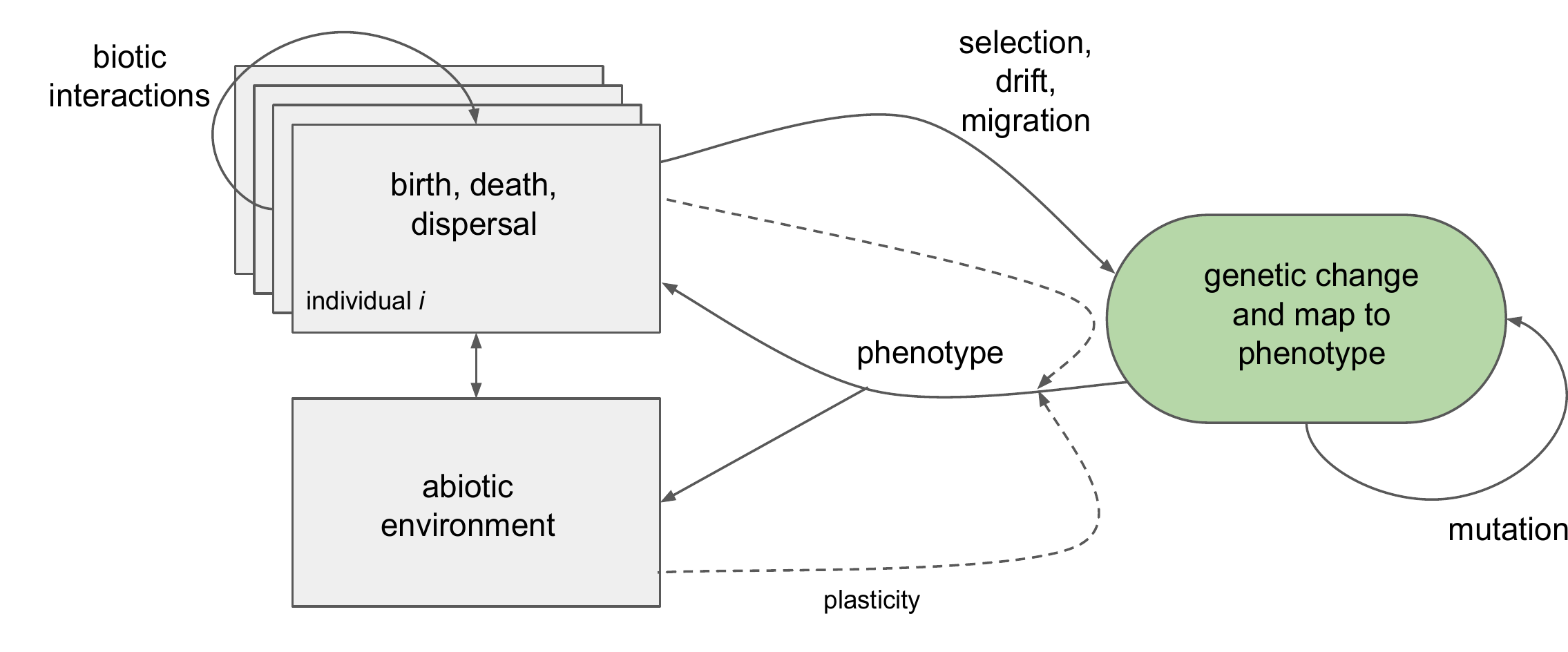}
	\caption{Mechanistic underpinnings of EEFs. Ecological dynamics (left) are driven by individual-level properties (birth, death, dispersal). Interactions between individuals of the same or different species (biotic interactions) impact these properties, which may lead to density-dependence, for example. Individuals interact with the abiotic environment and vice versa. Importantly, these ecological settings will impact selection, drift and migration (eco-to-evo). Evolution is governed by the interaction between these processes, genetic constraints and mutations. The resulting phenotype is subsequently determined by the genotype-phenotype map. Ultimately, the phenotype will impact ecology (evo-to-eco) by changing births, death, dispersal and the abiotic environment. Plasticity (dashed lines) may modulate the phenotype  and, hence, the dual effects of the organism on biotic and abiotic environments.}
	\label{Fig:discussion}
\end{figure}

Because different formalisms originate from different fields, they often rely on differing assumptions. For instance, the time scales on which processes occur and the sources of genetic variation are important consideration of the different modelling formalisms \citep{Lande2007, Saether2015}. This has made some formalisms more focussed on analysing evolutionary end-points and long-term dynamics (AD), while others have focused on short-term dynamics from one generation to the next (QG). However, in both formalisms incorporating EEFs is feasible. The separation of time scales also means that the form of the feedback may change when we move from one dynamical regime to the other, which has been well studied in host-parasite models \citep{Lenski1994, Day2007, Gandon2009, Lion2018}. However, most interest probably lies in predicting the mid-term dynamics of an EEF system. To approach this properly, an important issue for future theoretical work will be to develop mechanistic models for the dynamics of phenotypic and genotypic variation in populations evolving at this mid-term time scale of tens to hundreds of generations (see Fig.~\ref{Fig:discussion} for an individual based perspective). This would reveal for instance whether EEFs are time dependent and how common they are expected to be.

Overall, recent models have become more elaborate. However, increased complexity and realism often trades off with generalism. For example, IBMs might be able to capture more complex and realistic situations, however, they often lack generality and it may be difficult to determine the mechanisms underlying a certain result. As a consequence these studies must provide additional tests that either involve simulations where the presumed feedback is absent, or provide a simplified analytical model \citep[e.g.,][]{Kubisch2016, Branco2018}.

The challenge today certainly consists in pursuing new, more integrative and mechanistic modelling avenues which have the potential to include different aspects of generalism, such as genotype-phenotype mapping, plasticity, population and spatial structure (Fig.~\ref{Fig:discussion}). Current theory has greatly increased our understanding of EEFs \citep{McPeek2017}, but these feedbacks have been primarily explored within hierarchical levels of ecosystem organization, be they spatial or temporal hierarchies, and have often involved only single or a few independently evolving traits. While the presence of a hierarchical organization of ecosystems is well established \citep{Melian2018}, it is an ongoing challenge to identify the relevant hierarchical levels and their interdependencies to understand EEFs. 

Currently, the leading graphical model adopts an implicit hierarchy with feedbacks between levels from genes, to traits, to populations, to communities, to ecosystem processes \citep[][see also Fig.~\ref{Fig:introduction}A for a simplification]{Hendry2017}. Making such a conceptual model more mechanistic requires understanding how interactions at one scale (gene regulatory networks or complex traits) affect processes at different scales (trait-dependent species interactions). One such modelling attempt by \citet{Melian2018} links ecological and evolutionary networks in a meta-ecosystem model, taking into account demography, trait evolution, gene flow, and the ecological dynamics of natural selection. Such process-based models can yield new insights into the mechanistic basis of EEFs in more complex natural scenarios. For example, some of the most important processes are summarized in Fig.~\ref{Fig:discussion} which expands the conceptual model presented in Fig.~\ref{Fig:introduction}A to a more mechanistic level. With this representation we propose that feedbacks are best conceptualized as emerging from individual-level interactions \citep[see also][]{Rueffler2006a}, with dispersal and interactions with the abiotic environment leading to the emergence of the relevant hierarchical complexity.

To date, there is still limited integration of empirical data, either from natural or experimental settings, with theoretical models \citep[but see for instance][]{Fischer2014a, Fronhofer2015b, Huang2017, deMeesterinprep}. Connecting theory to controlled laboratory or field experiments will allow to experimentally assess the feedback. Most importantly, generating alternative models and predictions based on alternative hypotheses and confronting these with data will make inferences stronger. Clearly, making these connections between theory and empirical data is more difficult when studying ecology and evolution in the wild \citep{Hendryinprep} and ecological pleiotropy may even cancel out EEFs \citep{DeLong2017}. We suggest that a three-way interaction between theory, laboratory-based experiments and comparative data from natural communities may be most productive. Understanding how prevalent EEFs are in nature is critical for links to policy makers. In this context it is central to know whether species coexistence can be better predicted with feedbacks, whether the evolution of resistance may be faster with feedbacks, or whether population size can be better controlled when including feedbacks, to name but a few examples.

Understanding the dynamical consequences of EEF is more important than ever in a rapidly changing world. Theoretical models are the best avenue to create testable hypotheses, however, isolation among subdisciplines and methods leads to confusion, reduced inference and will not advance the field. We suggest that taking a mechanistic, individual-based perspective \citep{Rueffler2006a} as outlined in Fig.~\ref{Fig:discussion} can be productive for developing novel and synthetic theory.


\section*{Acknowledgements}
The authors would like to thank the editors of the special issue for bringing us together.

\section*{Author's contributions}
\begin{itemize}
\itemsep-1.5em	
\item Conceived study: Lynn Govaert, Emanuel A. Fronhofer, Blake Matthews\\
\item Led study: Lynn Govaert, Emanuel A. Fronhofer\\
\item Performed literature search and analysis: all authors\\
\item Drafted the manuscript: Lynn Govaert, Emanuel A. Fronhofer, Blake Matthews\\
\item Contributed to revisions: all authors\\
\end{itemize}

\section*{Data accessibility}
All papers used for this review are cited in the text.


\end{document}